\documentclass[prb, preprint, amsmath,amssymb, showpacs, longbibliography, superscriptaddress]{revtex4-1}
\usepackage{graphicx} 
\usepackage{dcolumn}  
\usepackage{bm}       
\usepackage{amsfonts}
\usepackage{dsfont,easyReview}
\usepackage{color}

\usepackage{soul,xcolor}
\usepackage{ulem}
\usepackage{color}

\begin{document}

\preprint{}
\title{Optical Control of Ultrafast Photocurrent in Graphene}
\author{Navdeep Rana}
\affiliation{%
Department of Physics, Indian Institute of Technology Bombay,
            Powai, Mumbai 400076  India}

\author{Gopal Dixit}
\email{gdixit@phy.iitb.ac.in}
\affiliation{%
Department of Physics, Indian Institute of Technology Bombay,
            Powai, Mumbai 400076  India}
\affiliation{%
Max-Born Institut, Max-Born Stra{\ss}e 2A, 12489 Berlin, Germany }
\date{\today}



\begin{abstract}
The ability to manipulate electrons with the intense laser pulse  enables an unprecedented control 
over the electronic motion on its intrinsic timescale. 
Present work explores the desired control of photocurrent generation in monolayer graphene on ultrafast timescale. 
The origin of photocurrent is attributed to the asymmetric residual electronic population  
in the conduction band after the end of the laser pulse, which also facilitates valley polarization.
Present study offers a comprehensive analysis of the differences between these two observables, namely 
photocurrent and valley polarization.  
It is found that the corotating circularly polarized $\omega-2\omega$ laser pulses allow the generation of photocurrent but no valley polarization, whereas counterrotating circularly polarized $\omega-2\omega$ laser pulses yield significant 
valley polarization without any photocurrent in graphene. 
Different laser parameters, such as subcycle phase, wavelength, and 
intensity provide different knobs to control the generation of the photocurrent.  
In addition, threefold increase in the photocurrent's amplitude can be achieved by altering electronic properties of  graphene via strain engineering. 
Our findings reveal intriguing underlying mechanisms 
into the interplay between the symmetries of the graphene's electronic structure 
and the driving laser pulses, shedding light on the potential for harnessing graphene's properties for novel applications in ultrafast photonics, optoelectronic devices, and quantum technologies.  

\end{abstract}

\maketitle

\section{Introduction} 
Thanks to  stupendous advancements in laser technology and synthesis of novel materials 
that have allowed us to investigate  laser-driven ultrafast phenomena in a variety of solids~\cite{tielrooij2015generation, yoshioka2022ultrafast, rees2020helicity, wang2015ultrafast,  sirica2021shaking, kruchinin2018colloquium, ghimire2019, weber2021ultrafast}. 
In particular, recent years have witnessed an upsurge in interrogating  laser-driven coherent 
electron dynamics in solids on attosecond timescale~\cite{bharti2022high, mrudul2020high, heide2021electronic, bharti2024photocurrent}.
In this regard,  various  static and dynamic symmetries play a key role in determining the response of electrons, 
within the solid under interrogation, to an externally applied perturbation.
The interaction of an intense laser pulse with a 
solid has  potential to alter the solid's symmetry, consequently revealing novel effects that were absent in solid at equilibrium conditions~\cite{lindner2011floquet,hu2020dynamical,rana2022high,neufeld2019floquet, nagai2020dynamical}.
Generally the absence of an inversion symmetry is a prerequisite to observe the photovoltaic effect and 
an emergence of the electric polarization in solids~\cite{ma2023photocurrent,grinberg2013perovskite,tan2016shift}.
However, few-cycle intense laser pulses are employed to induce photovoltaic effect in 
inversion-symmetric solids via controlling either  carrier-envelope phase 
(CEP)~\cite{fortier2004carrier,roos2005characterization,paasch2014solid,rybka2016sub,heide2021optical,boolakee2022light} or  the chirp~\cite{wu2020waveform} of the laser waveform. 
Recent work has  demonstrated that the combined effect of the CEP and chirp allows us to control 
photocurrent in two-dimensional material~\cite{zhang2023residual}.

The inherent asymmetry of the CEP-stabilized few-cycle laser pulses results asymmetric residual electronic population in the conduction band, which allows photocurrent generation.
Alternatively, combination of the two-color relatively ``long" pulses,  
instead of the monochromatic CEP-stabilized few-cycle pulses, is employed to realize asymmetric residual population in the conduction band within the Brillouin zone. 
Corotating circularly-polarised  $\omega-2\omega$ laser pulses are used to generate photocurrent in a variety of solids~\cite{neufeld2021light}.
In addition, it has been demonstrated that the counterrotating circularly polarized  $\omega-2\omega$  
laser pulses are able to generate photocurrent in the inversion-symmetric topological materials~\cite{ikeda2023photocurrent}. 
Note that the counterrotating $\omega-2\omega$  combination has also been used to induce an asymmetric population, which allowed us to realize valley polarization in an inversion-symmetric two-dimensional material~\cite{mrudul2021light}.

Laser-driven photocurrent in a solid  is defined as
$ \mathbf{J} = \int [\rho_{n}(\mathbf{k}) - \rho_{n}(-\mathbf{k})] \frac{\partial \mathcal{E}_{n}}{\partial \mathbf{k}} d\mathbf{k}$, 
 where 
$\rho_{n}(\mathbf{k}) $ is the residual electronic population in the 
$n^{\textrm{th}}$-conduction band after the laser pulse, and $\mathcal{E}_{n}(\mathbf{k})$ 
is the energy dispersion of the solid as a function of crystal momentum $\mathbf{k}$~\cite{soifer2019band, bharti2023tailor}.  
In addition, the population difference is related to the valley polarization  as
$\eta = 2 (n_{\mathbf{K}}-n_{\mathbf{K}^{\prime}})/(n_{\mathbf{K}}+n_{\mathbf{K}^{\prime}})$ 
with $n_{\mathbf{K}(\mathbf{K}^{\prime})} = \int_{0}^{\mathbf{K}(\mathbf{K}^{\prime})}\rho_{n}(\mathbf{k})d\mathbf{k}$~\cite{mrudul2021light}.  
From the expressions of $ \mathbf{J}$ and $\eta$, it appears 
that both the quantities rely on the asymmetric distribution of the residual  electronic population. 
However, the mechanism through which the asymmetry of a laser pulse results in valley polarization or the photocurrent 
is not obvious  {\it a priori}. 
Moreover,   the importance lies not only in the generation of photocurrent but also  in the imperative need to tailor and 
enhance photocurrent's amplitude for numerous technological applications~\cite{Pusch_2023, koppens2014photodetectors, xia2009ultrafast, paasch2014solid}. 
From the definition of the photocurrent, it is straightforward  that the 
photocurrent's amplitude can be controlled by two ways: 
either by increasing the asymmetry in the residual  electronic population, i.e., $ [\rho_{n}(\mathbf{k}) - \rho_{n}(-\mathbf{k})]$, or by engineering a solid in such a  way that the topology of the energy dispersion, i.e., $\partial \mathcal{E}_{n} / \partial \mathbf{k}$, can be tailored.  

\begin{figure}
\includegraphics[width=0.8\linewidth]{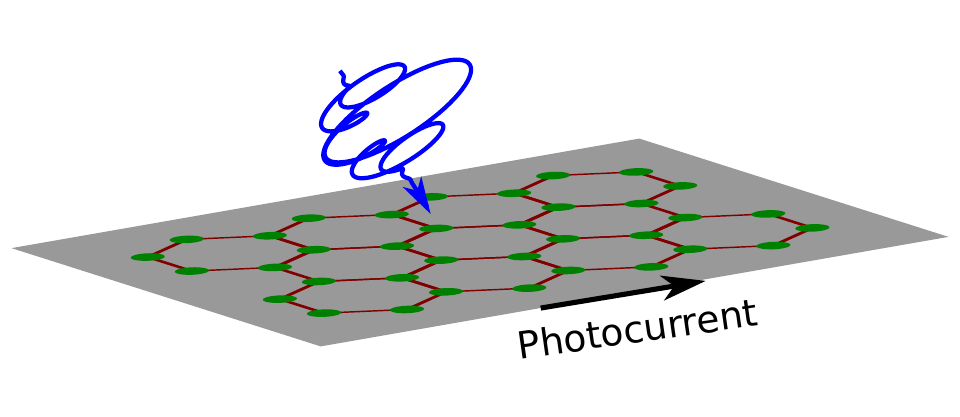}
\caption{Schematic of the photocurrent generation in a two-dimensional monolayer graphene via tailored  $\omega-2\omega$ laser pulses.}
\label{schematic}
\end{figure}

Present work is dedicated  to comprehensive investigation about the underlying mechanism for the photocurrent generation and valley polarization  in the monolayer graphene via tailored $\omega-2\omega$ laser pulses as illustrated in Fig.~\ref{schematic}. 
In the following, we will explore the conditions to generate photocurrent and/or valley polarization.  
Additionally,  we  will illustrate the subcycle phase between the $\omega-2\omega$ laser pulses allow us to tailor photocurrent. 
To gain deeper insights on enhancing the asymmetry in the residual population, we systematically explore 
how different laser parameters, such as intensity and wavelength of the $\omega-2\omega$ laser pulses impact
photocurrent generation. 
Our findings reveal a remarkable threefold enhancement  in the photocurrent's amplitude   
when pristine graphene undergoes strain engineering, underscoring its potential 
for tailoring photocurrent generation significantly.
To generate perturbative photocurrent  in graphene, coherent interference of $\omega$ and $2\omega$ interband excitation pathways was theoretically predicted ~\cite{mele2000coherent, rioux2011current}  and experimentally realized~\cite{glazov2014high, sun2010coherent}. 
Moreover, conditions to obtain asymmetric population in the conduction band to generate perturbative photocurrent for linearly polarized, corotating and counterrotating circularly polarized lasers were  discussed.

\section{Theoretical Method}
The nearest-neighbor tight-binding model is employed to describe monolayer graphene with
the corresponding Hamiltonian  as~\cite{reich2002tight}
\begin{equation}
\mathcal{H}_\textbf{TB} = - t_{0}\sum_{i\in nn} ~e^{i\textbf{k}\cdot \textbf{d}_i} \hat{a}_\textbf{k}^{\dagger} \hat{b}_\textbf{k} + \textrm{H.c.} 
\label{eq:tb}
\end{equation}
Here, $t_0$ = 2.7 eV represents the hopping energy,  
$\textbf{d}_i$ corresponds to the separation between the atom and its nearest neighbor, 
such that $|\textbf{d}_i| = a$ = 1.42 \AA~signifying the inter-atomic distance and $\textrm{H.c.} $ stands for Hermitian conjugate. 
The operator  $\hat{a}_\textbf{k}^{\dagger}$ ($\hat{b}_\textbf{k}$) denotes the creation (annihilation) operators for 
 A (B) type atom in the unit cell.
Graphene has zero band gap  at the two points in the Brillouin zone known as $\mathbf{K}$ and $\mathbf{K}^{\prime}$ points around which energy dispersion is linear in nature. 
Interaction of the tailored laser pulses with graphene is simulated by employing density-matrix-based  
semiconductor Bloch equation formalism within the Houston basis~\cite{mrudul2021high, rana2023all}.
The total time-dependent current as a function of crystal momentum is calculated as
\begin{equation}
\textbf{J}(\textbf{k}, t)   = \sum_{m,n  \in \{c,v\} } \rho_{mn} ^{\textbf{k}} (t)  \textbf{p}_{nm} ^{\textbf{k}+\mathbf{A}(t)}.
\end{equation}
Here, $\textbf{p}_{nm}(\textbf{k}) = \langle n,\textbf{k}|\nabla_\textbf{\textbf{k}}\mathcal{H}_\textbf{TB}| m,\textbf{k}\rangle$ is the momentum matrix element with $| n,\textbf{k}\rangle$ and $| m,\textbf{k}\rangle$ as the eigen-states of 
$\mathcal{H}_\textbf{TB}$ and $\mathbf{A}(t)$ is the  vector potential of the driving laser pulse.  
By performing the integration with respect to $\textbf{k}$ over the entire Brillouin zone,  we obtain the total time-dependent  current $\textbf{J}(t)$.
The total vector potential corresponding to the circularly polarized $\omega-2\omega$ laser pulses is defined as
\begin{equation}
    \mathbf{A}(t) = \dfrac{A_{0} f(t)}{\sqrt{2}} \biggl(
    \Bigl[ \cos(\omega t + \phi) + \cos(2\omega t) \Bigr]\hat{\mathbf{e}}_{x}+ \Bigl[ \sin(\omega t + \phi) \pm \sin(2\omega t) \Bigr]\hat{\mathbf{e}}_{y} 
    \biggr).
\end{equation}
Here, + (-) represents the corotating (counterrotating) configuration, 
$A_{0}$ is the amplitude of the vector potential, $f(t)$ is the temporal envelope for the driving field, 
and $\phi$ is subcycle phase between $\omega$ and $2\omega$ pulses. 
In the present  work,  the fundamental  $\omega$ pulse has a  wavelength of 2.0 $\mu$m 
with peak intensity of 0.3 TW/cm$^2$ and  
10 fs is used to mimic the decoherence between electron and hole during strong-field driven dynamics~\cite{ranasymmetry}. 
The intensity ratio of the $\omega$ and $2 \omega$ pulses are unity throughout this work. 
The driving laser pulse has eight cycles with a sine-square envelope. 
The laser parameters used in this work are below the damage threshold of graphene~\cite{currie2011quantifying}, 
and similar laser parameters have been used to investigate  strong-field driven electron dynamics in graphene~\cite{heide2018coherent,higuchi2017light}.  

\section{Results and Discussion}
\begin{figure}
 \includegraphics[width= 0.8\linewidth]{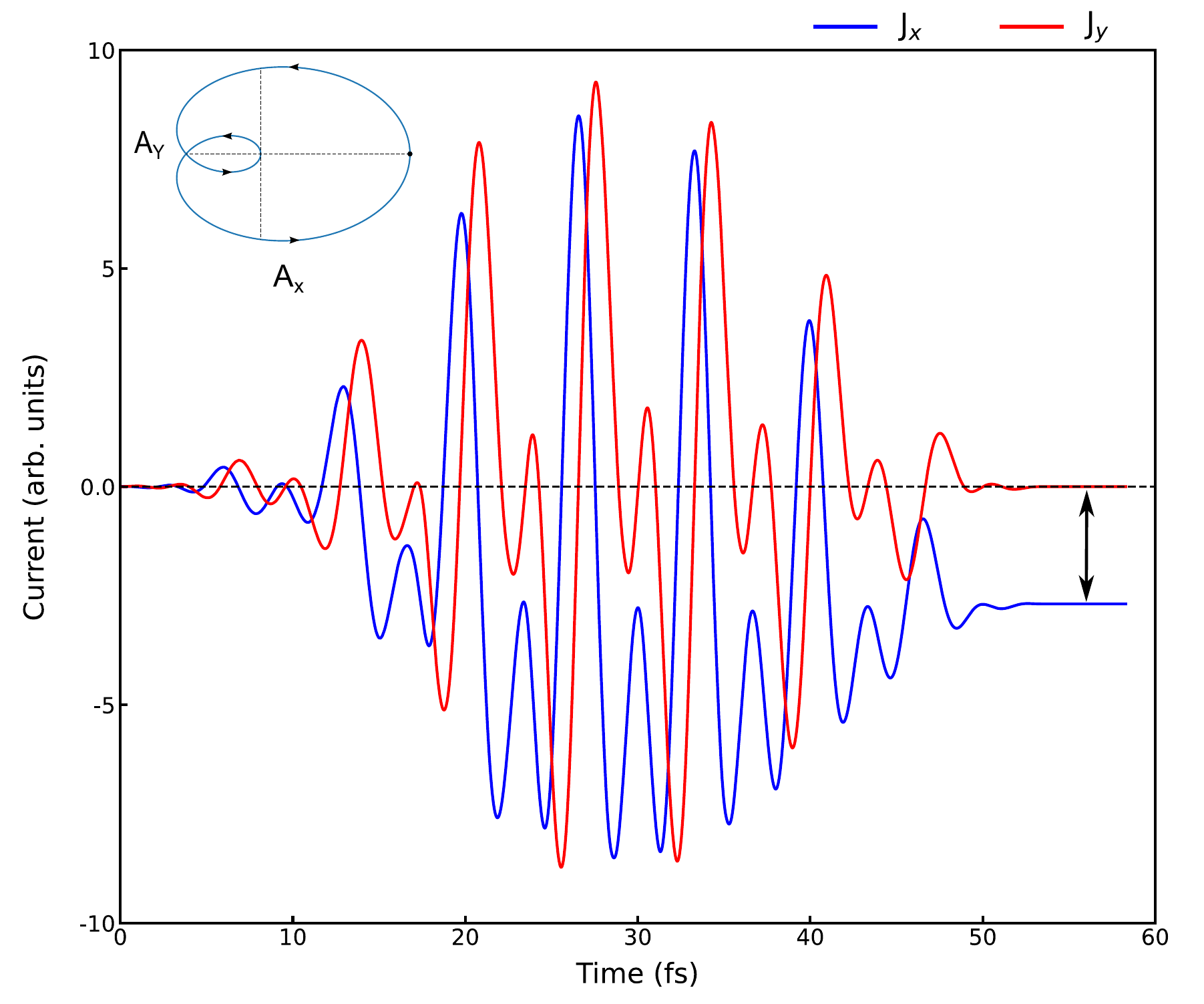}
 \caption{Time-dependent total current in a monolayer graphene driven by corotating $\omega-2\omega$ laser pulses with  zero subcycle phase  for wavelength  of 2 $\mu$m and intensity of 0.3 TW/cm$^{2}$. 
The Lissajous curve corresponding to the total vector potential is shown in the inset.}
 \label{photocurrent}
\end{figure}

The total  current in graphene induced by the corotating $\omega-2\omega$ laser pulses  
with zero subcycle phase is shown in Fig.~\ref{photocurrent}. 
The  photocurrent is nonzero  along the $x$ direction only with negative in magnitude, which is consistent with the recent report~\cite{neufeld2021light}.
The reason for the nonzero photocurrent can be attributed to the axial symmetry with respect to the $x$ axis of the 
the vector potential as reflected from the inset of Fig.~\ref{photocurrent}. 
Consequently, any current generated along the positive and negative $y$ axes cancels each other, resulting  zero  photocurrent along the $y$ direction. 
However, it is crucial to emphasize that the vector potential is asymmetric about the $y$ axis, which  
leads to an asymmetric residual electronic population with respect to the $y$ axis, 
and therefore the observed photocurrent is along $x$ direction.
At this juncture, it is pertinent to wonder how the generated photocurrent varies with the subcycle phase of the  corotating $\omega - 2\omega$ pulses.  

\begin{figure}
\includegraphics[width=0.8\linewidth]{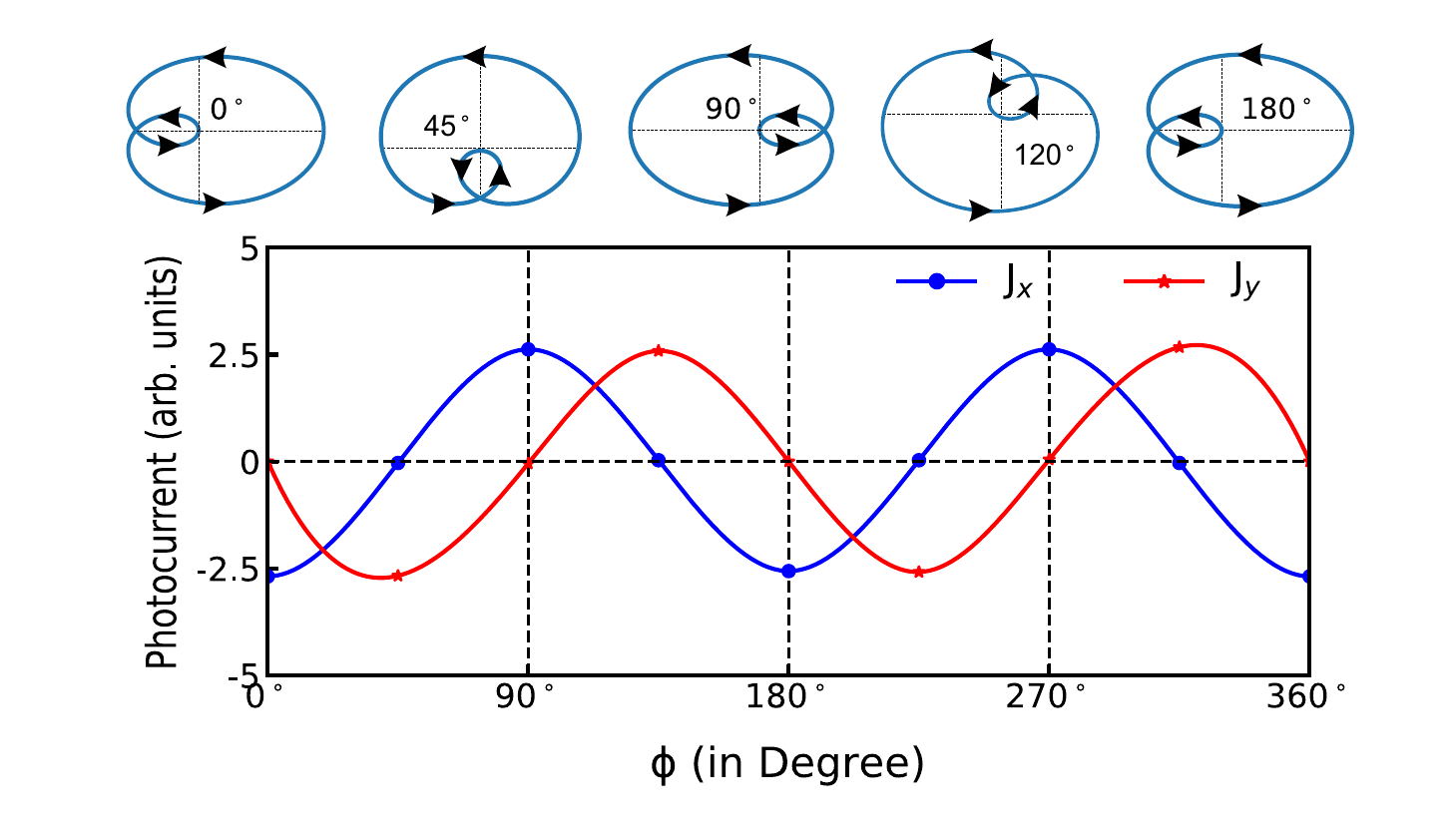}
\caption{Sensitivity of the photocurrent with respect to the subcycle phase of the corotating $\omega-2\omega$ laser pulses. 
Lissajous curves of the total vector potential for different subcycle phases  
of  the $\omega-2\omega$ laser pulses are depicted in the top panel. 
Rest of the laser parameters are the same as in Fig.~\ref{photocurrent}.}
\label{photocurrent_control}
\end{figure}

Figure~\ref{photocurrent_control} presents  the sensitivity of the photocurrent with the subcycle phase of the corotating  pulses.
It is apparent that the directionality of the photocurrent can be strategically tailored via controlling  the subcycle phase.  
Photocurrent becomes zero along the $x$ direction while becoming finite 
along the $y$ direction for the subcycle phase value of 45$^\circ$ as evident from Fig.~\ref{photocurrent_control}. 
On the other hand, the photocurrent is positive and finite along the $x$ direction but becomes zero along the $y$ direction for the subcycle phase of 90$^\circ$. 
For this value of the phase, the total vector potential manifests asymmetry with respect to the $x$ axis, while 
 symmetric along the $y$ axis as reflected from the Lissajous curve at the top panel. 
Consequently, this configuration results finite photocurrent  along the $x$ direction only. 
Moreover,  generation of the positive photocurrent can be attributed to the laser waveform of the vector potential, 
which predominantly  drives more electrons towards the positive $k_{x}$ direction in the momentum space.
On the other hand, when the subcycle phase acquires a value of 120$^\circ$, the vector potential exhibits  
asymmetry along both the $x$  and $y$ axes, resulting in finite photocurrent along both axes.
From the analysis of the Lissajous curve of the vector potential for different subcycle phases shown in the top panel of Fig.~\ref{photocurrent_control}, it can be concluded that 
the directionality of the generated photocurrent is inherently linked to the asymmetry present in the vector potential of the corotating laser pulses. 
Moreover, the intensity ratio between $\omega$ and $2 \omega$ pulses does not alter the symmetry of the Lissajous curve as shown earlier~\cite{rana2022corotating}.
Thus, the subcycle phase of the corotating $\omega-2\omega$  pulses provides a tunable  knob to tailor the photocurrent in a controlled manner. 

The generation of the photocurrent relies on the fact that $ [\rho_{n}(\mathbf{k}) - \rho_{n}(-\mathbf{k})] \neq 0$. 
Along the same line, it is natural  to think that the asymmetric residual population also allows us to observe  
valley polarization as the asymmetric characteristic  is intertwined with  the valley polarization as
$\eta = 2 (n_{\mathbf{K}}-n_{\mathbf{K}^{\prime}})/(n_{\mathbf{K}}+n_{\mathbf{K}^{\prime}})$ with 
$n_{\mathbf{K} (\mathbf{K}^{\prime})} = \int_{0}^{\mathbf{K} (\mathbf{K}^{\prime})}\rho_{n}(\mathbf{k})d\mathbf{k}$. 
In addition, it has been shown that the counterrotating $\omega-2\omega$ laser pulses lead to significant 
valley polarization in monolayer graphene~\cite{mrudul2021light,mrudul2021controlling}. 
Thus, it is crucial to delve deeper to unravel the underlying mechanisms  of  how corotating and counterrotating $\omega-2\omega$ laser pulses facilitate photocurrent  and  valley polarization in  graphene. 

\begin{figure}
\includegraphics[width= 0.8\linewidth]{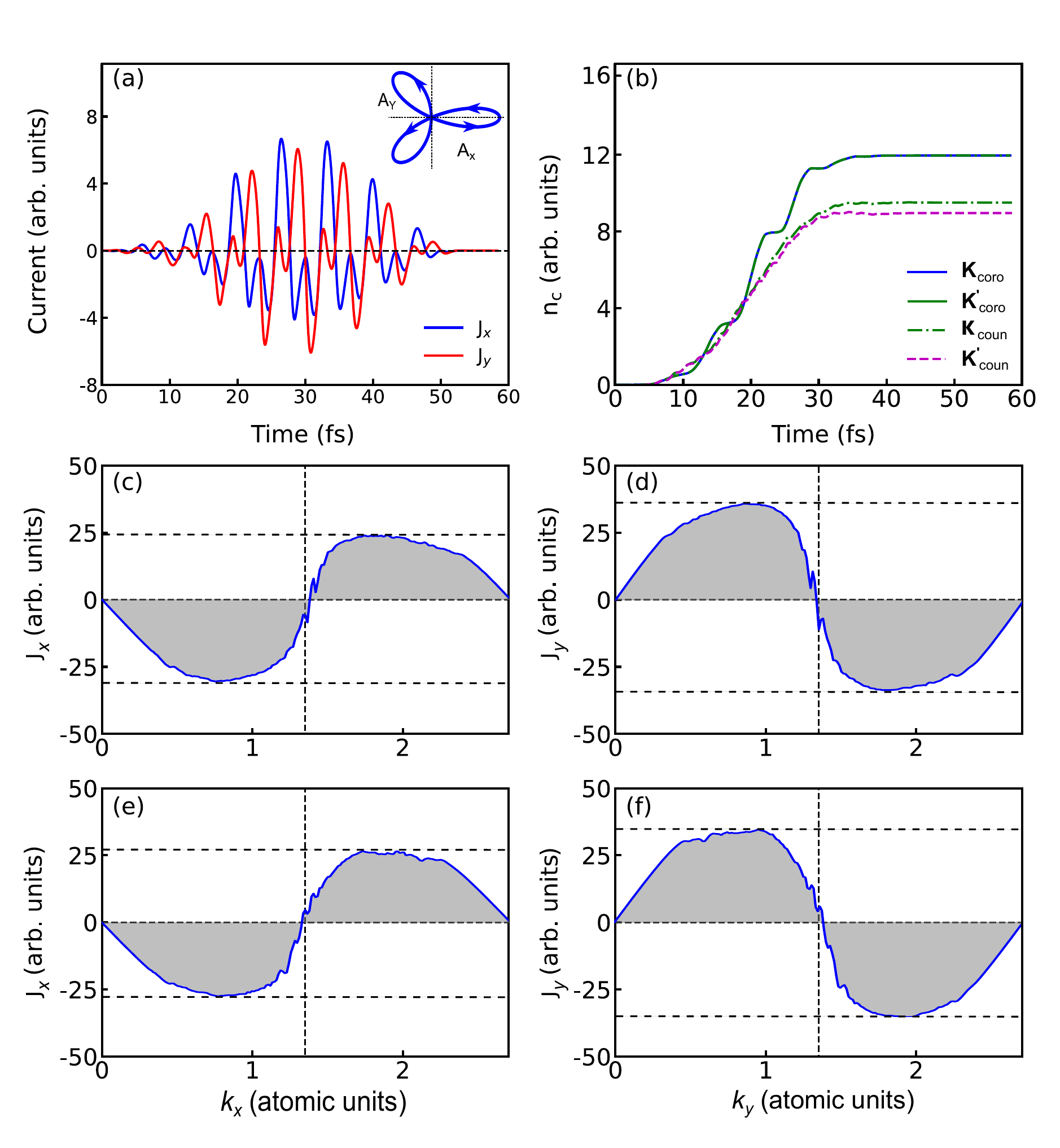}
\caption{(a) Total time-dependent current in graphene generated by the counterrotating $\omega-2\omega$ laser pulses with zero subcycle phase. The corresponding Lissajous curve to the total vector potential is shown in the inset.
(b) Time-dependent charge dynamics induced by corotating and  counterrotating configurations around 
$\mathbf{K}$ and $\mathbf{K}^{\prime}$ valleys. 
The Green dashed dot and violet dashed line correspond to the  charge dynamics for $\mathbf{K}$ ($\mathbf{K}_{\textrm{coun}}$) and $\mathbf{K}^{\prime}$ ($\mathbf{K}^{\prime}_{\textrm{coun}}$) during a counterrotating laser fields. 
On the other hand, blue and green solid lines stand for $\mathbf{K}$ ($\mathbf{K}_{\textrm{coro}}$) and $\mathbf{K}^{\prime}$ ($\mathbf{K}^{\prime}_{\textrm{coro}}$) associated with  the corotating configuration. 
The integrated photocurrent along the (c) $k_{x}$ and (d) $k_{y}$ axes for the corotating configuration. 
(e) and (f) are same as (c) and (d) for the counterrotating configuration, respectively.
The laser parameters are the same as in Fig.~\ref{photocurrent}.}
 \label{valley}
\end{figure}

There is no photocurrent  either along the $x$ or $y$ direction when graphene is irradiated by  
counterrotating $\omega-2\omega$ laser  pulses as shown in  Fig.~\ref{valley}(a). 
An absence of the photocurrent can be attributed to the underlying symmetry of the total vector potential associated with the counterrotating laser fields. 
For a given subcycle phase of the  counterrotating  field, the vector potential exhibits asymmetry with respect to the $y$ axis~\cite{rana2022corotating}. 
However, if one takes the projection of the vector potential along the $x$ and $y$ axes, it is straightforward to see that the vector potential is symmetric about both the  $x$ and  $y$ axes, which results in no photocurrent. 
However, it is essential to emphasis that this asymmetric residual population in the momentum space also  has  implications for the valley polarization.

The integrated residual populations in the vicinity of $\mathbf{K}$ and $\mathbf{K}^{\prime}$ valleys, induced by corotating and
counterrotating $\omega-2\omega$ laser pulses, are illustrated in Fig.~\ref{valley}(b).
It is interesting to observe that  $n_{\mathbf{K}}$ and $n_{\mathbf{K}^{\prime}}$ are identical for the 
corotating fields,  which results in no valley polarization. 
On the other hand, $n_{\mathbf{K}}$ and $n_{\mathbf{K}^{\prime}}$ are significantly different in the case of  the 
counterrotating fields, which provide significant valley polarization. 
Analysis of Fig.~\ref{valley}(b) for counterrotating configuration  is consistent 
with previous reports~\cite{mrudul2021light,mrudul2021controlling}.  
Moreover, corotating configuration leads to 
photocurrent but no valley polarization, whereas counterrotating 
configuration results in  valley polarization with zero photocurrent. 

To gain a deeper understanding of these intriguing behaviors, 
let us analyze the integrated photocurrent for both the configurations. 
As previously discussed, the vector potential of the corotating configuration exhibits asymmetry along the $y$ axis for the  phase of 0$^{\circ}$. Consequently, the photocurrent integrated along the $k_{y}$ axis also displays asymmetry as reflected  from  Fig.~\ref{valley}(c).
On the other hand, the photocurrent integrated along the $k_{x}$ axis exhibits symmetry as depicted in 
Fig.~\ref{valley}(d), which allows  
photocurrent exclusively along the $x$-direction in this scenario and no  photocurrent along the $y$ direction.
In the case of the counterrotating configuration, the  photocurrent integrated along both the $k_{y}$ and $k_{x}$ axes exhibit symmetry as shown in Figs.~\ref{valley}(e) and \ref{valley}(f), respectively. 
Thus there is no photocurrent in either direction for the counterrotating configuration.
For significant valley polarization, it is important to note that the total vector potential must align with the trigonal warping of the graphene's energy band structure 
around $\mathbf{K}$ and $\mathbf{K}^{\prime}$ valleys to induce asymmetry in the population distribution within these valleys~\cite{mrudul2021light}.  
In the case of the corotating configuration, the total vector potential does not align with the band structure around   $\mathbf{K}$ or $\mathbf{K}^{\prime}$ valley, leading to no valley polarization. 
However, counterrotating configuration exhibits  trefoil-shaped Lissajous structure [see inset of Fig.~\ref{valley}(a)], 
which aligns perfectly 
with one valley over the other, giving rise to the observed valley asymmetry as reflected  in Fig.~\ref{valley}(b).

\begin{figure}
\includegraphics[width=0.8\linewidth]{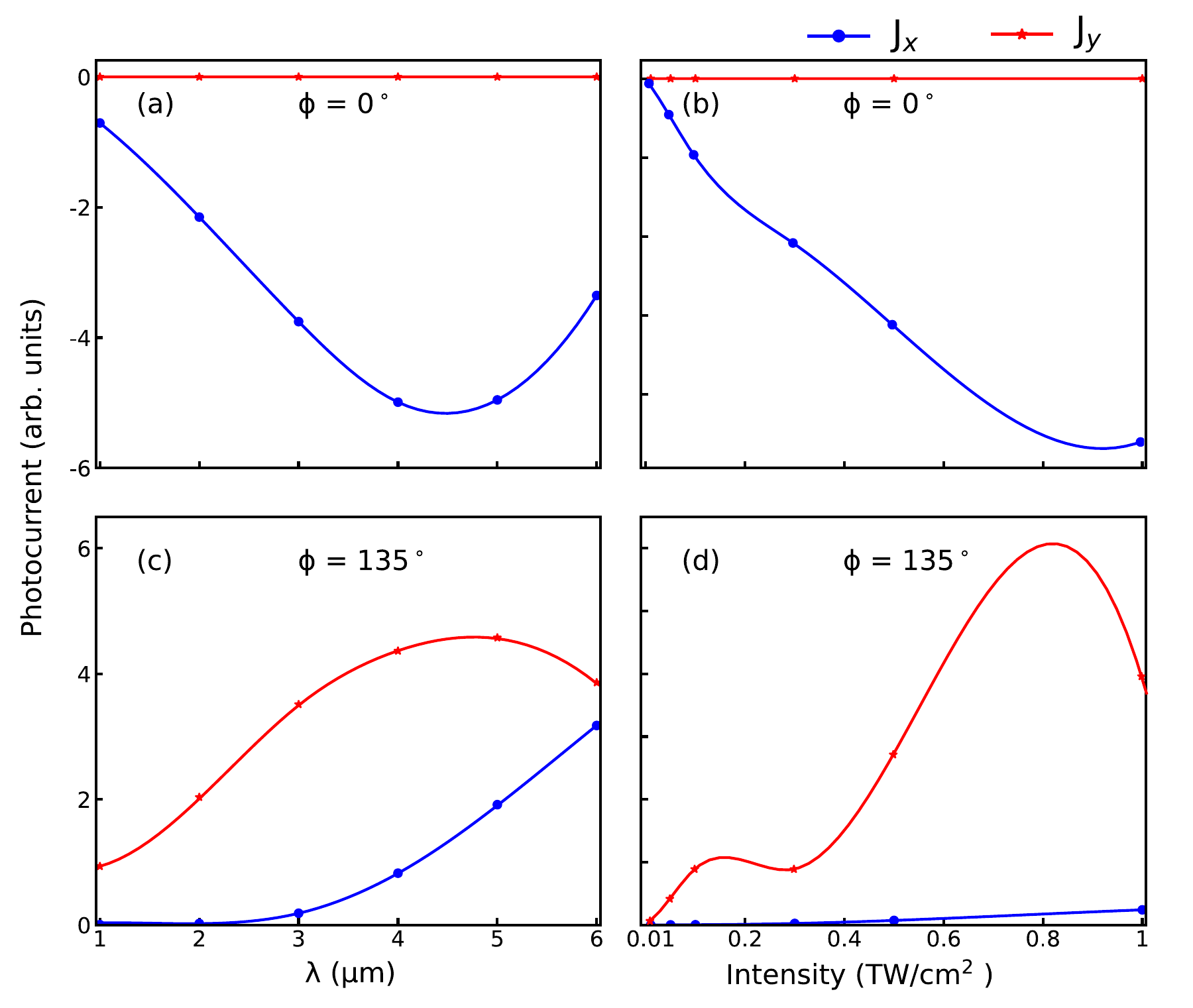}
\caption{Variation of the photocurrent's amplitude, $\mathsf{J}_{x}$ and $\mathsf{J}_{y}$, 
with respect to wavelength of the fundamental $\omega$ pulse of the 
corotating $\omega-2\omega$ pulses for the subcycle phase (a) 0$^\circ$ and (c) 135$^\circ$. 
(b) and (d) are same as (a) and (c) for the intensity of the $\omega$ pulse, respectively.}
\label{wavelength_response}
\end{figure}

So far we have investigated how the subcycle phase impacts the directionality and amplitude of the photocurrent driven 
by corotating $\omega-2\omega$ pulses.
Note that the magnitude of the  photocurrent is similar for all the subcycle phase values.
Now let us turn our discussion to understand how the wavelength and intensity of the driving laser influence the photocurrent's amplitude.  
The variation in the photocurrent's amplitude with respect to the wavelength of the $\omega$ pulse 
is demonstrated in Figs.~\ref{wavelength_response}(a) and ~\ref{wavelength_response}(c) for the subcycle phase 
of 0$^\circ$ and 135$^\circ$, respectively. 
The photocurrent exhibits a wavelength-dependent behavior along the $x$ direction ($\mathsf{J}_{x}$), 
which is increasing with wavelength initially  and starts decreasing beyond a critical value of the wavelength as illustrated in Fig.~\ref{wavelength_response}(a). 
Note that the photocurrent along the $y$ direction ($\mathsf{J}_{y}$) is zero for $\phi = 0^\circ$ as shown in Fig.~\ref{photocurrent_control}.
It is known that the electrons are driven along the direction of the laser's  electric field, 
compelling them to move away from the $\mathbf{K}$ and $\mathbf{K}^{\prime}$ points in the Brillouin zone~\cite{mrudul2021light}. 
As the wavelength increases, electrons are pushed farther into the Brillouin zone, introducing greater asymmetry and consequently causing an increase in the photocurrent. 
However, beyond a specific wavelength threshold, electrons start traversing between the $\mathbf{K}$ and $\mathbf{K}^{\prime}$ regimes in the Brillouin zone. 
This introduces interference between these paths, leading to a reduction in the asymmetry in the residual population,
which results  in a decrease in the photocurrent as depicted in Fig.~\ref{wavelength_response}(a).

As discussed earlier, the vector potential exhibits asymmetry along both the $x$ and $y$ axes for $\phi = 135^\circ$, 
resulting photocurrent in both directions as evident in Figs.~\ref{photocurrent_control}  and \ref{wavelength_response}(c). 
It is noteworthy that $\mathsf{J}_{x}$ becomes finite for a given threshold of the wavelength.  
Beyond this threshold, $\mathsf{J}_{x}$  increases as the wavelength increases.
This peculiar  behavior of the photocurrent arises due to the limited asymmetry of the total vector potential with respect to the $y$-axis. 
Hence, the asymmetry of the vector potential along the  $y$ axis becomes noticeable only when electrons are driven significantly  away from the $\mathbf{K}$ and $\mathbf{K}^{\prime}$ points within the Brillouin zone.
This observed threshold behavior in the $x$ direction  provides  insights into how the magnitude of the photocurrent responds to variations in wavelength, shedding light on the intricate interplay between the photocurrent and laser pulse characteristics.
The variation of $\mathsf{J}_{y}$ with wavelength in this case is similar to the one for $\phi = 0^\circ$  with the similar underlying mechanism.  

How the intensity of the laser pulse influences the photocurrent is illustrated  in 
Figs.~\ref{wavelength_response}(b) and ~\ref{wavelength_response}(d) for  $\phi = $ 0$^\circ$ and 135$^\circ$, respectively. 
Initially the photocurrent increases with the intensity.  
However, as the intensity approaches around  TW/cm$^{2}$, the photocurrent starts reducing  due to a highly nonlinear effect~\cite{neufeld2021light}.
In addition, the decrease in the  photocurrent's amplitude can be attributed to the fact that  beyond a certain threshold of the laser's intensity,  the photocurrent  reverses  its direction as discussed previously in Refs.~\cite{zhang2022bidirectional,higuchi2017light}.

Throughout our exploration so far, we have delved into regulating photocurrent in pristine graphene by controlling  the subcycle phase as well as tuning  
wavelength or intensity of corotating $\omega-2\omega$ laser pulses. 
Nevertheless, it is also feasible to tailor the graphene's properties 
by engineering  to control the photocurrent.
Strain engineering has emerged as a highly effective approach for manipulating the characteristics of graphene. Furthermore, it is important to note that the strain naturally arises when graphene is placed  on a substrate,  
such as hexagonal boron nitride during device fabrication. 
In the following, we will  focus on the most common form of strain engineering in graphene, i.e., uniaxial strain.

When the uniaxial strain is present in graphene,  atomic positions and the lattice vectors are transformed as 
\begin{equation}
\tilde{\mathbf{u}} =  (\mathbf{I} + \mathcal{\mathbf{\varepsilon}}) \cdot \mathbf{u},
\label{eq:transform}
\end{equation}
where \textbf{I} is the 2$\times$2 identity matrix and \textbf{$\varepsilon$} is the strain tensor.
The displacement of the atoms results in atomic  distance alteration in graphene, which results in modification 
to the hopping energy [see Eq.~(\ref{eq:tb})]. 
The  modified hopping energy is described by an exponential function of the interatomic distance defined as $\gamma_{mn} = \gamma_{0} e^{-(\textbf{d}_{mn}-a)/b}$ with 
$\textbf{d}_{mn}$ as the separation vector between nearest-neighbor atoms in graphene and  
$b$ = 0.32$a$ as the decay distance~\cite{pereira2009tight}. 
The change in the hopping term  is  updated into the tight-binding Hamiltonian for the strained graphene.

\begin{figure}
\includegraphics[width=\linewidth]{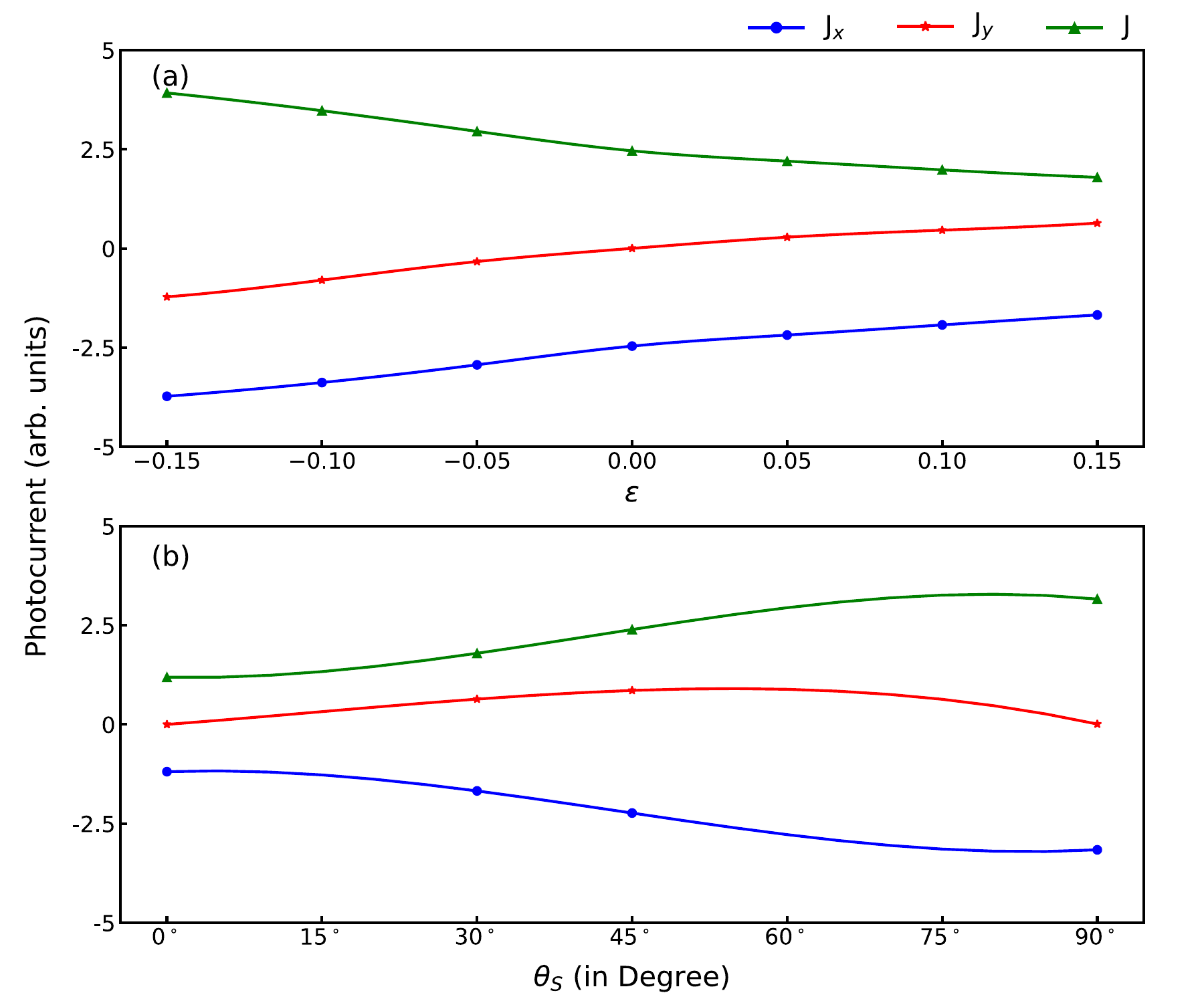}
\caption{Variation in total photocurrent ($\mathsf{J}$) and its components in strained graphene as a function of the (a)
strain's strength, $\varepsilon$, applied along $30^\circ$ with respect to the $k_{x}$ axis, and 
(b) strain angle, $\theta_{s}$, with  respect to the $k_{x}$ axis for $\varepsilon = 0.15$. 
Rest of the laser parameters are same as in Fig.~\ref{photocurrent}.}
\label{strain_vary} 
\end{figure}

Figure~\ref{strain_vary}(a) presents the variation in the photocurrent for different 
strengths of the strain ($\varepsilon$) applied along $30^\circ$ with respect to the $k_{x}$ axis. 
The photocurrent is decreasing as the strain's strength  is increasing and vice versa. 
In the absence of any strain, we have finite $\mathsf{J}_{x}$, which is 
related to the asymmetry of the vector potential of the corotating fields with $0^\circ$. 
However, the reflection symmetries along the $\mathsf{XZ}$ and $\mathsf{YZ}$ planes are broken when the uniaxial strain is along $30^\circ$. 
As a result of this symmetry breaking, photocurrent is also generated along the $y$ direction. 
Graphene is stretched along the direction of the strain and compressed along 
 the perpendicular direction to the applied strain when the uniaxial strain is tensile in nature, i.e., 
 strain's strength is positive. 
The direction of the photocurrent can be tuned from positive to negative in $y$-direction by
changing the nature of strain from tensile to compressive as evident from Fig.~\ref{strain_vary}(a). 
The extent of asymmetry increases as the strain's strength increases, which
results in an increase in the photocurrent along the $y$ direction.
However, photocurrent, along the $x$ direction, decreases as strain transits from compressive to tensile strain. 
The reduction in the photocurrent can be attributed to a decrease in the group velocity along the $x$ direction. Here, the group velocity is defined as $\mathbf{v}_{g}$ = $\partial \mathcal{E}(\mathbf{k}) / \partial \mathbf{k} $, which is proportional to intraband  current. 

Let us turn our discussion to when the direction of the applied strain, $\theta_{s}$, 
is varied for a given value of $\varepsilon$. 
The sensitivity of the photocurrent  with respect $\theta_{s}$ for  $\varepsilon =$ 0.15 is presented in 
Fig.~\ref{strain_vary}(b).   
$\mathsf{J}$ monotonically increases with $\theta_{s}$, whereas 
$\mathsf{J}_{y}$ increases and reaches to maximum at $\theta_{s} = 45^{\circ}$.
After that, $\mathsf{J}_{y}$ starts reducing and reaches to zero at $\theta_{s} = 90^{\circ}$. 
For $\theta_{s} = 0^{\circ}$, 
graphene is distorted in such a way that the reflection symmetries along 
$\mathsf{XZ}$ and $\mathsf{YZ}$ planes are preserved, which 
allows the photocurrent along the  $x$ direction only. 
The reflection plane symmetries along both the planes are broken as 
$\theta_{s}$ changes from 0$^{\circ}$ to 45$^{\circ}$.
Moreover, the extent of the asymmetry increases with $\theta_{s}$ and reaches to  the maximum at  45$^{\circ}$, which
allows $\mathsf{J}_{y}$ to be maximum at that angle. 
On the other hand, extent of the asymmetry reduces as $\theta_{s}$ changes from  45$^{\circ}$ to 90$^{\circ}$. 
For $\theta_{s} = 90^{\circ}$, the reflection symmetries are preserved, which leads $\mathsf{J}_{y}$ to be zero.  
An increase in $\mathsf{J}_{x}$  can be  
attributed to an increase in the group velocity  along the $x$ direction with $\theta_{s}$. 
Thus, analysis of Fig.~\ref{strain_vary} establishes that the strain engineering is viable route to control the photocurrent in graphene. 
Moreover, photocurrent's amplitude and direction can be controlled by 
applying appropriate strain with suitable strength along the desired direction.   

In summary, we have explored how the waveform of the tailored laser pulses allows us to create  asymmetric residual electronic population, which in turn generates photocurrent and/or valley polarization in  graphene. 
The detailed underlying mechanisms responsible for generating photocurrent and/or observing valley polarization, and their intricate connection with the symmetries of the corotating and counterrotating circularly-polarized 
$\omega-2\omega$ laser pulses, and energy dispersion of graphene are analysed in depth. 
We have also explored how the photocurrent can be tailored by tuning the laser parameters of the 
corotating  laser, such as subcycle phase, intensity and wavelength. 
In addition, we have illustrated that the photocurrent's amplitude can be also significantly boosted  by manipulating the electronic structure of  the monolayer  graphene via uniaxial strain engineering.
Present research offers an avenue for graphene-based ultrafast photodetectors and photonics devices operating on ultrafast timescale.

\section*{Acknowledgments}
We acknowledge fruitful discussions with M. S. Mrudul (Uppsala University, Sweden) and Amar Bharti (IIT Bombay). G.D. acknowledges financial support from SERB India (Project No. MTR/2021/000138). 

\newpage
%

\end{document}